\documentclass[groupedaddress,showpacs,showkeys,amssymb,eqsecnum,aps,superscriptaddress]{revtex4}

\usepackage{lmodern}

\usepackage[]{graphicx}
\usepackage[]{graphics}
\usepackage{amsmath}
\usepackage{epsf}                                                                                           
\usepackage{color}                   
\usepackage{verbatim}
\usepackage[mathscr]{euscript}

\usepackage{hyperref}

\newcommand{\be}{\begin{equation}}

\newcommand{\ee}{\end{equation}}

\begin{document}

 \author{F. T. Brandt}  
 \email{fbrandt@usp.br}
 \affiliation{Instituto de F\'{\i}sica, Universidade de S\~ao Paulo, S\~ao Paulo, SP 05508-090, Brazil}

\author{J. Frenkel}
\email{jfrenkel@if.usp.br}
\affiliation{Instituto de F\'{\i}sica, Universidade de S\~ao Paulo, S\~ao Paulo, SP 05508-090, Brazil}

 \author{D. G. C. McKeon}
 \email{dgmckeo2@uwo.ca}
 \affiliation{
 Department of Applied Mathematics, The University of Western Ontario, London, Ontario N6A 5B7, Canada}
 \affiliation{Department of Mathematics and Computer Science, Algoma University, Sault Ste.~Marie, Ontario P6A 2G4, Canada}

\author{G.  S.  S.  Sakoda}   
\email{gustavo.sakoda@usp.br}
\affiliation{Instituto de F\'{\i}sica, Universidade de S\~ao Paulo, S\~ao Paulo, SP 05508-090, Brazil}
 

\title{On the analytically-improved running coupling in QCD}

\date{\today}

\begin{abstract}
We examine, in the `t Hooft renormalization scheme, the analytic running coupling $\bar\alpha_t(Q^2)$ in QCD, using the two-loop $\beta$-function with positive expansion parameters $\beta_0$  and $\beta_1$. An exact integral representation  is  derived for this causal coupling, which is fully expressed in terms of the imaginary part of the Lambert function $W$. This integral form manifestly accounts for the universal value of the infrared limit
$\bar\alpha_t(Q^2=0)= 4 \pi /\beta_0$.
\end{abstract}


\maketitle

\section{Introduction}\label{sec1}

In perturbative calculations of physical processes, there appear at higher orders ultraviolet divergences, which require a procedure (called the renormalization scheme) for their subtraction.
At finite order in perturbation theory, a number of ambiguities occur as the renormalization procedure is not uniquely specified and, in particular,
amplitudes calculated in perturbation theory appear to depend on an arbitrary renormalization scale (usually denoted by $\mu$).
However, since these amplitudes are connected to physical observables, these cannot depend on such a parameter
(or any other parameter that characterize the renormalization scheme).
This can be achieved if the coupling becomes a function of $\mu$, which leads to the independence of observables on the choice of this scale parameter. The renormalization group (RG) is closely related to the scale invariance of physical systems.
Furthermore, it 
improves the properties of the perturbative series in the ultraviolet region, by summing the radiative corrections that occur in higher orders of perturbation theory \cite{r1,r2,r3,r4,r5,Lavrov:2012xz}. Due to asymptotic freedom \cite{r6}, physical quantities in QCD at high momentum squared $Q^2$ can be evaluated by solving the RG equations in terms of a power series involving the perturbative running coupling  $\alpha(Q^2)$. 

It is well known that such a running coupling leads to unphysical singularities like the Landau pole \cite{r7}, which renders perturbation theory useless for low-energy processes in QCD. A resolution of this problem was proposed a long time ago \cite{r8}, when the RG method was unified with the requirement of analyticity. This led to the concept of an analytic running coupling $\bar\alpha(Q^2)$ that is free of unphysical singularities. Subsequently, other methods have been developed through the use of dispersion relations that reflect the causality principle \cite{r9,r10,r11,r12}, and successfully applied in QCD. The analytic approach preserves the ultraviolet behavior in the ultraviolet region, which ensures asymptotic freedom but leads to essential modifications of the behaviour in the infrared region.
There are no unphysical singularities in the low energy region.
An important feature of this approach is the presence of a universal value of the analytic coupling $\bar\alpha(Q^2 = 0) = 4 \pi/\beta_0$, where $\beta_0$ is the coefficient of the $\beta$-function at one loop.
Several arguments for the universality of the analytic running coupling have been previously given in in the literature, from various points of view \cite{r9,r10,r11,r12,r13,r14,r15,r16,r17,Milton:1997miMilton:1998jy,r19aa}.
Another relevant property of the analytic approach is that it allows us to define the causal running coupling $\bar\alpha(Q^2)$ both  in the space-like as well as in the time-like domains, in a consistent way. These features enable meaningful calculations of hadronic processes in QCD, like the inelastic lepton-hadron scattering and the electron-positron annihilation into hadrons as described, for example, in the reviews \cite{Prosperi:2006hx,Deur:2016tte} and the references cited therein.

Moreover, there are also other analytic methods for improving the QCD perturbation theory \cite{Deur:2016tte}.
These generally involve some criterion for making a choice of the renormalization scale $\mu$, or other parameters that characterize a renormalization scheme, so that the perturbative result best approximates the exact (scheme-independent) result.
In this context, we mention the ``Optimized Perturbation Theory'', which consists in improving the convergence of the perturbative expansion by choosing the value of $\mu$ according to a criterion of minimum sensitivity \cite{Stevenson:1980du,Stevenson:1981vj}. We also point out the ``Fastest Apparent Convergence'' technique \cite{Grunberg:1980ja,Grunberg:1982fw}, that  amounts  to selecting the scale $\mu$ so that the next-to leading and higher order coefficients of the perturbative series are set to zero. This optimization procedure is related to effective charges 
\cite{Grunberg:1980ja,Grunberg:1982fw,Brodsky:2011zza,Brodsky:2011ig}, which 
consist in defining new running couplings that are more directly connected with physical observables. In addition, we  mention the ``Principle of  Maximum Conformality'' \cite{Deur:2016tte} which improves the QCD predictive power by removing the  dependence of physical predictions on the choice of the renormalization scheme. There has also been much work towards unifying the different approaches,
including the AdS-CFT/Dyson-Schwinger methods \cite{deTeramond:2008ht,Cui:2019dwv}, 
which may lead to a consistent universal running coupling in QCD.

The purpose of this letter is to give an exact integral representation for the causal $\bar\alpha(Q^2)$ coupling in the 't Hooft scheme, which leads in a simple and direct way to the universal value of its infrared limit. In section \ref{sec2} we succinctly review some features of RG equations and of the `t Hooft renormalization scheme \cite{r18} where the $\beta$-function has exactly the two-loop form, which will be useful subsequently.
This approach removes 
the arbitrariness that occurs in other renormalization schemes.
%
%
Here we also give the known form of the perturbative running coupling at two-loops in terms of the Lambert function \cite{r19}. In section \ref{sec3} we derive an exact integral form (Eq. \eqref{eq21}) for the analytic running coupling, which is entirely  expressed in terms of the imaginary part of Lambert's function. A useful analytic approximation of this integral form that is good to within 
$7.7$\%  of the exact expression is given in Eq. \eqref{eq316}.   
We conclude this note with a summary of the results in section \ref{sec4}, where we also briefly discuss an application of this approach to the $e^+ e^-$ annihilation into hadrons.
A more systematic analysis of the role of the 't Hooft coupling, when considering the renormalization scheme dependence in perturbative QCD, is given in Appendix \ref{appC}.

\section{The perturbative running coupling}\label{sec2}
The perturbative running coupling is defined by the RG differential equation 
\be\label{eq1}
Q^2 \frac{d}{d Q^2} \alpha(Q^2) = -\beta(\alpha(Q^2)); \;\;\; \alpha(\mu^2)=\alpha ,
\ee
where $\mu$ is the renormalization point, $\alpha=g^2/4\pi$ and $g$ is the renormalized coupling constant.

The $\beta$-function may be written in the form
\be\label{eq2}
\beta(\alpha(Q^2)) = \frac{\beta_0}{4\pi}\alpha^{2}(Q^2) + \frac{\beta_1}{(4\pi)^2}\alpha^{3}(Q^2)+\frac{\beta_2}{(4\pi)^3}\alpha^{4}(Q^2)+\cdots ,
\ee
where, up to two-loop order ($N_f$ being the number of active quark flavors)
\be\label{eq3}
\beta_0 = 11 - 2 N_f/3; \;\;\;\beta_1 = 102 - 38 N_f/3 .
\ee

G. `t Hooft showed \cite{r18} that it is possible to choose a coupling $\alpha_t$ such that the $\beta$-function has just the 2-loop form,
 which is invariant under changes of the renormalization scheme.
This condition leads to the relation \cite{r20,r21}
\be\label{eq4}
\alpha = {\alpha_t}+\left(\frac{1}{4\pi}\right)^2
\frac{\beta_2}{\beta_0} \alpha_t^3 + \cdots .
\ee
It has been argued \cite{r20,r22,r23} that an all-order summation of the terms which depend explicitly on $\mu$ and the expansion parameters $\beta_2,\dots$, yields  a $\beta$-function that  is consistent with  `t Hooft scheme.

At two loops, a straightforward integration of Eq. \eqref{eq1} gives \cite{r9,r10,r11,r12} 
\be\label{eq5}
\ln\left(\frac{Q^2}{\Lambda^2}\right)=\frac{1}{a(Q^2/\Lambda^2)}-\frac{\beta_1}{\beta_0^2}\ln\left(1+\frac{\beta_0^2}{\beta_1}\frac{1}{a(Q^2/\Lambda^2)}\right),
\ee
where $a(Q^2/\Lambda^2) \equiv (\beta_0/4\pi) \, \alpha(Q^2/\Lambda^2)$ and $\Lambda$ is the QCD scale parameter, 
which is associated with the $\alpha(1)=\infty$
boundary condition in Eq. \eqref{eq1}.

An inversion of Eq. \eqref{eq5} can be written in terms of the Lambert multi-valued function $W$ \cite{r19}, which is defined as
\be\label{eq6}
W(z) \exp[W(z)] = z
\ee
where
\be\label{eq7}
z(Q^2) = -\frac{1}{e}\left(\frac{Q^2}{\Lambda^2}\right)^{-\beta^2_0/\beta_1} 
\ee
($e$ is the Euler number).  One can now verify that an exact solution of the Eq. \eqref{eq5} is given by \cite{r15,r17}
\be\label{eq8}
a(z) = -\frac{\beta_0^2}{\beta_1} \frac{1}{1+W(z)}.
\ee
We note that in the `t Hooft scheme, this expression would give the complete solution. The requirement that $a(Q^2)$ is real and positive for positive $Q^2$ and that it should vanish in the limit $Q^2\rightarrow\infty$, determines the appropriate branch of the Lambert function. In this work we consider QCD with  $N_f \le 6$, which implies that $\beta_0$ and $\beta_1$ in Eq. \eqref{eq3} are positive. In this case, the function $z(Q^2)$ in Eq. \eqref{eq7}  is negative and the correct physical branch that ensures asymptotic freedom is $W_{-1}(z)$, which becomes negative and infinite in the limit $z\rightarrow 0$.

\section{The analytic running coupling}\label{sec3}
This causal coupling may be constructed through a dispersion relation as \cite{r9,r10,r11,r12} 
\be\label{eq9}
\bar a_t(Q^2/\Lambda^2) \equiv\frac{1}{\pi}\int_0^\infty d\sigma \frac{\Im a_t(-\sigma)}{\sigma+Q^2},
\ee
where 
$Q^2>0$ for space-like momentum transfer.


To one-loop order, the perturbative running coupling has the form
\be\label{eq10}
a_t^{(1)}(Q^2/\Lambda^2)=\frac{1}{\ln(Q^2/\Lambda^2)}.
\ee
Performing the analytical continuation $\sigma\rightarrow\sigma+i\epsilon$ and evaluating the imaginary part of $a_t(-\sigma)$ in  Eq. \eqref{eq9}, yields the one-loop analytic running coupling function
\be\label{eq11}
\bar a_t^{(1)}(Q^2/\Lambda^2)=
\int_0^\infty d\sigma\frac{1}{\sigma+Q^2}\frac{1}{\ln^2(\sigma/\Lambda^2)+\pi^2}
=\frac{1}{\ln(Q^2/\Lambda^2)}+\frac{1}{1-Q^2/\Lambda^2}.
\ee
The first term is the usual perturbative running coupling at one-loop. The second term ensures the correct analytic properties, by cancelling the Landau pole present in the first term. The above expression has a physical cut when the real part of $Q^2$ is negative, and no other singularities, so that it is consistent with causality. We also note  that $\bar a_t^{(1)}(0)=1$.

We next proceed to the two-loop case, by using the expression given in Eq. \eqref{eq8} for the perturbative running coupling. To this end, one needs to evaluate the imaginary part of this coupling. One then gets in Eq. \eqref{eq9} a complicated integrand, because the integration variable $\sigma$ occurs implicitly in the Lambert function $W$. For this reason, it is convenient  to transform this equation  into  an equivalent integral, where the integration variable is just the imaginary part of $W$. To this end, we will proceed in two steps as follows. We first change the integration variable to $z$, as given by Eq. \eqref{eq7}, with $Q^2$ replaced by $\sigma$
\be\label{eq12}
z(\sigma) = -\frac{1}{e}\left(\frac{\sigma}{\Lambda^2}\right)^{-\beta_0^2/\beta_1}
\ee
so that the integration will occur along the negative real $z$-axis
\be\label{eq13}
\bar a_t(Q^2/\Lambda^2) = -\frac{e}{\pi} \int_{-\infty}^0 dz \frac{
  (-e z)^{-\beta_1/\beta_0^2-1}}{(-e z)^{-\beta_1/\beta_0^2}+Q^2/\Lambda^2}\Im\frac{1}{1+W(z(-\sigma))}.
\ee

We now pass to calculate the imaginary part of $1/[1+W(z(-\sigma))]$, using Eq. \eqref{eq12} and making the analytic continuation $\sigma\rightarrow\sigma+i\epsilon$. In this way, we find that we need to evaluate the Lambert function $W$ at the point  $Z(\sigma)=z(\sigma)\exp(i\pi\beta_0^2/\beta_1)$. Thus, we write
\be\label{eq14}
Z = x + i y
\ee
and
\be\label{eq15}
W = \tau + i \eta .
\ee
Using the above equations together with the relation $W(Z) \exp[W(Z)] = Z$, and equating the imaginary parts of this relation, one finds that
\be\label{eq16}
x = z(\sigma) \cos(\pi \beta_0^2/\beta_1) = e^{\tau}(\tau \cos\eta-\eta\sin\eta)
\ee
and
\be\label{eq17}
y = z(\sigma) \sin(\pi \beta_0^2/\beta_1) = e^{\tau}(\eta \cos\eta+\tau\sin\eta).
\ee
Dividing the last two equations and using a trigonometric identity, one obtains 
\be\label{eq18}
\tau=-\eta\cot(\eta - \pi\beta_0^2/\beta_1).
\ee
Substituting this relation in Eq. \eqref{eq16} and using  a trigonometric identity, we get 
\be\label{eq19}
z(\sigma(\eta)) = -\frac{\eta}{\sin(\eta-\pi\beta_0^2/\beta_1)}
\exp-(\eta\cot(\eta-\pi\beta_0^2/\beta_1)) .
\ee

We can now evaluate the imaginary contribution  in Eq. \eqref{eq13}, by using the relations \eqref{eq8}, \eqref{eq12}, (\ref{eq14}-- \ref{eq18}), obtaining 
\be\label{eq20}
\Im\frac{1}{1+W(z)}=\Im\frac{1}{1-\eta\cot(\eta-\pi\beta_0^2/\beta_1)+i\eta}
=-\frac{\eta}{1-2\eta\cot(\eta-\pi\beta_0^2/\beta_1)+\eta^2/\sin^2(\eta-\pi\beta_0^2/\beta_1)}.
\ee
Substituting this relation in Eq. \eqref{eq13} and changing the variable of integration from $z(\eta)$ to $\eta$ by using Eq. \eqref{eq19}, one obtains after a straightforward calculation, the following integral form for the analytic running coupling 
\be\label{eq21}
\bar a_t(Q^2/\Lambda^2) = \frac{1}{\pi}\int_{\pi\beta_0^2/\beta_1}^{\pi+\pi\beta_0^2/\beta_1} d\eta
\frac{1}
{1 + (Q^2/\Lambda^2)[-ez(\eta)]^{\beta_1/\beta_0^2}}.
\ee


The above result expresses the causal coupling $\bar a(Q^2/\Lambda^2)$ in terms of an integral involving the imaginary part $\eta$ of the Lambert function, which corresponds to the branch $W_1$ \cite{r19}. This exact integral representation directly shows that, at the point $Q^2= 0$, this coupling equals to $1$, which implies that $\bar\alpha_t(Q^2=0)=4\pi/\beta_0$, in accordance with the universal value obtained by other methods. 

\begin{figure}[th]
\includegraphics[scale=0.55]{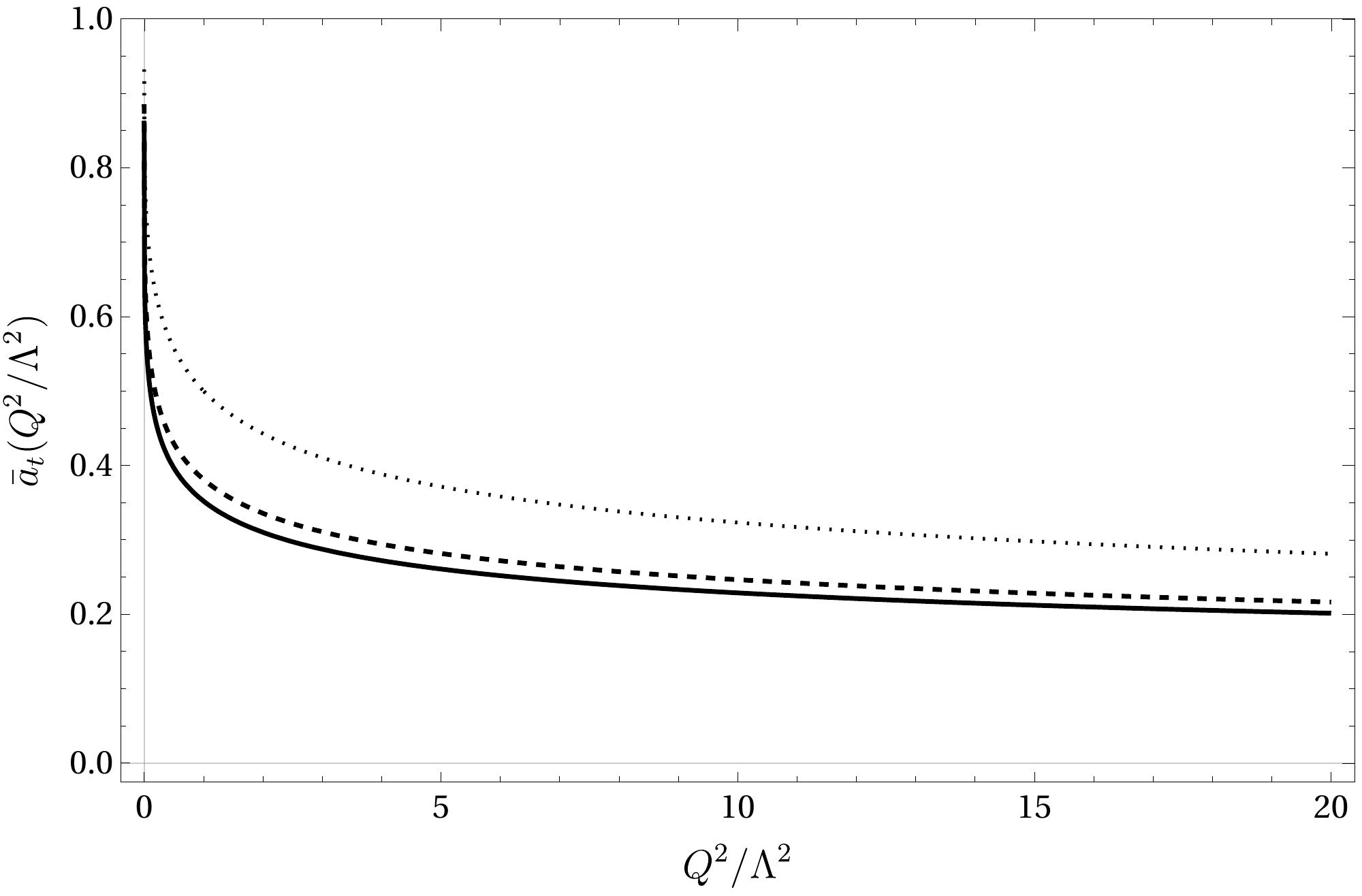}
  \caption{The behaviour of the integral in Eq. \eqref{eq21} for $\beta_1/\beta_0^2=64/81$ (solid) and $26/49$ (dashed) corresponding respectively to $N_f=3$ and $N_f=6$. The analytic one-loop result is plotted as a dotted line.}
\label{fig1}
\end{figure}

The above form of $\bar a_t(Q^2/\Lambda^2)$   
shows that it has a negative derivative whose magnitude 
decreases from infinity to zero as the value of $Q^2/\Lambda^2$ increases from zero to infinity.
Moreover, the second derivative of $\bar a_t(Q^2)$ is everywhere positive.
Thus, 
one may expect the integral to behave as depicted in Fig. \ref{fig1}, obtained by a numerical integration.
The graphs shown in Fig. \ref{fig1} illustrate the behaviour of this integral  for $\beta_1/\beta_0^2= 64/81$      and $\beta_1/\beta_0^2= 26/49$, corresponding respectively to $N_f = 3$ and
$N_f = 6$ in the Eq. \eqref{eq3}. One can see from this figure that for large values of $Q^2/\Lambda^2$, the analytic running coupling becomes, like the perturbative running coupling, of order  $1/\ln(Q^2/\Lambda^2)$. This may be understood by noticing  that in this case, the main contribution to the integral comes from the region where the integrand in 
Eq. \eqref{eq21} is of order $1$. But one can easily check that  this requires
values of $\eta$ near $\pi \beta_0^2/\beta_1$, in a range of order $1/\ln(Q^2/\Lambda^2)$. One then gets a slowly varying integrand, so the integral becomes proportional to the range of such a region. This explains the logarithmic form of the result obtained numerically for large values of $Q^2/\Lambda^2$.
The figure also shows that, in the infrared limit, there are no higher-loop 
corrections. 

The result shown in Fig. \ref{fig1} fully agrees with that obtained numerically at two loops, for $N_f=3$, in Ref. \cite{r17}.
It is also consistent with the result obtained in Ref. \cite{r19aa} by a different method, which does not involve the exact Lambert solution of Eq. \eqref{eq5}.
One may verify that in the limit $\beta_1\rightarrow 0$,
Eq. \eqref{eq21} reduces to Eq. \eqref{eq11}, as expected (see Appendix \ref{appA}).

Although the integration in Eq. \eqref{eq21} cannot be performed in closed form, a useful analytic approximation may be obtained by making in the dispersion relation \eqref{eq9} the shift $\sigma\rightarrow\sigma-Q^2$
so that $a_t(-\sigma)\rightarrow a_t(Q^2 - \sigma)$.
Then, in view of the fact that $a_t(Q^2-\sigma)$ is a slowly varying function of the logarithmic type, one may neglect here the $Q^2$ dependence.
This procedure is correct for small $Q^2$ and is also valid
for large values of $Q^2$, since in the ultraviolet 
region the imaginary part of $[a_t(-\sigma)-a_t(Q^2-\sigma)]$
leads to corrections of order $1/\ln^3(Q^2)$. In this way, we obtain the approximate analytic expression (see Appendix \ref{appB}) 
\be\label{eq316}
\bar a_t(Q^2/\Lambda^2) \simeq  1 - \frac{1}{\pi}\Im
\ln W_{1}[Z(Q^2/\Lambda^2)]
\ee
where
\be\label{eq317}
Z(Q^2/\Lambda^2) = -\frac{1}{e}(Q^2/\Lambda^2)^{-\beta_0^2/\beta_1}\exp(i \pi\beta_0^2/\beta_1). 
\ee
The above result has the correct limits both in the infrared as well as in the ultraviolet regions. Moreover, the numerical values displayed in table \ref{tab1} show that in the intermediate region this approximation is accurate to within $7.7\%$ (see also Ref. \cite{r17}).

\begin{table}[]
\begin{tabular}{|c|c|c|c|}
  \hline 
$Q^2/\Lambda^2$ & Eq. \eqref{eq21} & Eq. \eqref{eq316} & Relative difference   \\ 
  \hline
 0.00 & 1.0000 & 1.0000 & 0.000 \\
 1.00 & 0.3518 & 0.3411 & 0.030 \\
 10.0 & 0.2286 & 0.2112 & 0.076 \\
 20.0 & 0.2013 & 0.1858 & 0.077 \\
 40.0 & 0.1782 & 0.1651 & 0.074 \\
 80.0 & 0.1590 & 0.1482 & 0.068 \\
 100. & 0.1535 & 0.1435 & 0.066 \\
 200. & 0.1384 & 0.1303 & 0.059 \\
 400. & 0.1258 & 0.1193 & 0.052 \\
 800. & 0.1152 & 0.1100 & 0.045 \\
$10^6$ & 0.0616 & 0.0607 & 0.015 \\
$10^7$ & 0.0536 & 0.0531 & 0.009 \\
  \hline
\end{tabular}\caption{Numerical results from Eq. \eqref{eq21} and the approximation given by Eq. \eqref{eq316}, for $N_f=3$. The third column shows that the results agree to within 7.7\%. The last two lines show the level of accuracy of Eq. \eqref{eq316} for very large values of $Q^2/\Lambda^2$.}\label{tab1}
\end{table}

\section{Discussion}\label{sec4}

We have examined, in  the `t Hooft scheme, the behaviour of the analytic running coupling $\bar\alpha_t(Q^2)$ in QCD,  for  $\beta_0> 0$ and  $\beta_1> 0$. This coupling is independent of the renormalization procedure. This causal coupling, that is free from unphysical singularities, preserves the ultraviolet behaviour of the perturbative running coupling which ensures the asymptotic freedom of the theory, but modifies its behaviour in the infrared region. A relevant property of this approach is that the value of the analytic running coupling  at $Q^2=0$ is determined just by the one-loop corrections as $\bar\alpha_t(Q^2=0)=4\pi/\beta_0$. An exact integral representation of the analytic running coupling,
that is expressed in terms of the Lambert function,
is given in Eq. \eqref{eq21} which 
manifestly leads 
to this universal infrared limit.

We note here that the equation \eqref{eq5} may, alternatively, be solved by the iteration method \cite{r9,r10,r11,r12}. Although this procedure is accurate for large values of $Q^2/\Lambda^2$ when $a(Q^2/\Lambda^2)$ is small, it violates the analytical attributes of $a(Q^2/\Lambda^2)$ near the point $Q^2/\Lambda^2 = 1$ \cite{r17}. The present approach is consistent with the requirement of analyticity and thus is
convenient to investigate further the analytical properties of the causal coupling $\bar\alpha_t(Q^2/\Lambda^2)$.


These aspects may be useful to improve the calculation of physical quantities evaluated through the RG equations. 
For example, let us consider the ratio $R$ which is related to the total cross section for the  $e^+e^-$ annihilation into hadrons.
This is a function of the center of mass energy squared $s$ and of the renormalized coupling constant $\alpha$. At high energies, this ratio may be written in terms of a  perturbative series as  \cite{r24,r25} 
\be\label{eq24}
R(s/\mu^2,\alpha(\mu^2/\Lambda^2)) =
\sum_i Q_i^2\left[1+t_1(s/\mu^2)\alpha(\mu^2/\Lambda^2)
              +t_2(s/\mu^2)\alpha^2(\mu^2/\Lambda^2)+\cdots\right],
\ee
where $Q_i e$ is the charge of the $i$-th quark. The coefficients $t_1$, $t_2$ $\cdots$ include in general large logarithms like $\ln(s/\mu^2)$. The RG equation for this ratio implies that $R$ is independent of the renormalization point $\mu$, having  the form (see Appendix \ref{appC} )
\be\label{eq25}
R(s/\mu^2,\alpha(\mu^2/\Lambda^2)) = R(1,\alpha(s/\Lambda^2)),
\ee
where $\alpha(s)$ is the perturbative running coupling. One can thus express $R$ as follows
\be\label{eq26}
R = \sum_i Q_i^2 \left[1+t_1(1)\alpha(s/\Lambda^2)+t_2(1)\alpha^2(s/\Lambda^2)+\cdots\right].
\ee
When $\alpha(s/\Lambda^2)$ is given in the `t Hooft renormalization scheme, then $t_1$, $t_2$, etc, are renormalization scheme invariants \cite{r20}. The simplest way to improve the series for the ratio $R$ consists of replacing in the Eq. \eqref{eq26}, the perturbative running coupling by the analytic running coupling $\bar\alpha(s/\Lambda^2)$ defined in the time-like region.

An enhanced, but more involved, procedure has been elaborated in Refs. \cite{r12,Milton:1997miMilton:1998jy,r17,Prosperi:2006hx}.
It turns out that the time-like and the space-like
analytical running couplings 
may be related by the linear integral transformation ($q^2>0$)
\be\label{e44a}
         \bar\alpha(s)=\frac{i}{2\pi}
         \int_{s-i\epsilon}^{s+i\epsilon}\frac{dq^2}{q^2}\bar\alpha(-q^2),
\ee
where the contour integral is computed along a path in the analiticity region of the $\bar\alpha(-q^2)$ function.
These analytic running couplings have a common infrared stable point. In the ultraviolet limit, such couplings also have the same asymptotic behaviour. However, these functions generally differ in the intermediate energy range.

Relations between RG-invariant quantities defined in the Minkowski and Euclidean 
domains may be established by making use of linear integral transformations.
An example is the measurable ratio $R(s)$ which can be similarly related to 
the hadronic polarization function $\Pi(-q^2)$, calculable perturbatively in the Euclidean region.


A key feature of the analytic perturbation theory is the transformation of series involving powers of the space-like running coupling into functional expansions involving the time-like running coupling.
These expansion functions, which are free of unphysical singularities and have the universal value at the infrared stable point, yield an enhanced convergence of the analytic perturbation theory.  
Such a procedure has been performed for the $\tau$-lepton and $\Upsilon$ decays, for hadronic form factors as well as for
$e^+ e^-$  annihilation into hadrons, leading to results which are in a rather good agreement with the experimental observations \cite{r12,Prosperi:2006hx}.



\begin{acknowledgments}
  {We would like to thank CNPq (Brazil) for financial support.}
\end{acknowledgments}


\appendix

\section{The exact analytic coupling in the limit $\beta_1\rightarrow 0$}\label{appA}
From Eq. \eqref{eq21} one gets, in the limit $\beta_1\rightarrow 0$, the integral
($\phi = \eta - \pi \beta_0^2 /\beta_1 $)
\be\label{a1}
I(Q^2/\Lambda^2) = \frac{1}{\pi} \int_0^\pi d\phi \frac{1}{1+(Q^2/\Lambda^2)
  \exp( -\pi\cot\phi) }.
\ee
It is now convenient to make the change of variable $x=-\pi \cot\phi$ so that Eq. \eqref{a1} becomes
\be\label{a2}
I(Q^2/\Lambda^2) = \int_{-\infty}^\infty 
\frac{1}{1+(Q^2/\Lambda^2) e^x}\frac{dx}{x^2+\pi^2}.
\ee
Finally, making the change of variable $x=-\ln(\sigma/\Lambda^2)$, we obtain
\be\label{a4}
I(Q^2/\Lambda^2) = \int_{0}^\infty 
\frac{1}{\sigma+Q^2}\frac{d\sigma}{\ln^2(\sigma/\Lambda^2)+\pi^2},
\ee
which is equal to Eq. \eqref{eq11}. This is expected because in the limit
$\beta_1\rightarrow 0$, the exact analytic coupling should reduce to that obtained at one-loop. 

{Alternatively, using a  contour of integration in the complex plane which encloses only the poles inside the strip $0 < \Im z < 2\pi$, Eq. \eqref{a2} can be written as
(see Eq. (7.5.5) of \cite{ComplexVariables})
\be\label{a5}
I(Q^2/\Lambda^2) = \frac{1}{1-(Q^2/\Lambda^2)}
+ {\rm Res}_{z=z_0}\left[\frac{1}{1+(Q^2/\Lambda^2) e^z}\frac{1}{z-\pi i}
\right] = \frac{1}{1-(Q^2/\Lambda^2)} + \frac{1}{\ln(Q^2/\Lambda^2)},
\ee
where $z_0=-\ln(Q^2/\Lambda^2)+\pi i$ is the zero of ${1+(Q^2/\Lambda^2) e^z}$.}

\section{An approximate analytic result}\label{appB}
Making the shift $\sigma\rightarrow\sigma-Q^2$ in the dispersion relation \eqref{eq9} and neglecting the $Q^2$ dependence in the perturbative coupling $a_t$, we obtain
\be\label{b1}
\bar a_t(Q^2/\Lambda^2) \simeq \frac{1}{\pi} \Im \int_{Q^2}^\infty\frac{d\sigma}{\sigma}a_t(-\sigma).
\ee
Changing the variable of integration to $z(\sigma)$ given in Eq. \eqref{eq12} and using the relation \eqref{eq8}, one gets the integral
\be\label{b2}
\bar a_t(Q^2/\Lambda^2)\simeq \frac{1}{\pi} \Im \int_{Z(Q^2)}^{Z(\infty)}\frac{dZ}{Z}\frac{1}{1+W(Z)},
\ee
where $Z(\sigma) = z(\sigma)\exp(i\pi \beta_0^2/\beta_1)$.

Employing the identity \cite{r19}
\be\label{b3}
\frac{d W(Z)}{dZ} = \frac{1}{Z} \frac{W}{1+W}
\ee
one obtains from Eq. \eqref{b2} the following result
\be\label{b4}
\bar a_t(Q^2/\Lambda^2)
\simeq \frac{1}{\pi} \Im\int_{W_1[Z(Q^2)]}^{W_1[Z(\infty)]}\frac{dW}{W}
=1-\frac{1}{\pi} \Im\ln W_1[Z(Q^2/\Lambda^2)],
\ee
where we used the asymptotic form of the Lambert function $W_1$ \cite{r19},
which leads to the form given in Eq. \eqref{eq316}.

A simple example is provided by the one-loop analytic running coupling, in which case the Eq. \eqref{b1} yields
\be\label{b5}
\bar a_t^{(1)}(Q^2/\Lambda^2) \simeq \int_{Q^2}^\infty
\frac{d\sigma}{\sigma}\frac{1}{\ln^2(\sigma/\Lambda^2)+\pi^2}
=\frac{1}{2}-\frac{1}{\pi}\arctan\left[\frac{1}{\pi}\ln\left(\frac{Q^2}{\Lambda^2}\right)\right].
\ee
One may verify that at $Q^2 = 0$, Eq. \eqref{b5} gives a result equal to $1$, as expected.

For very large values of $Q^2/\Lambda^2$, one can expand Eq. \eqref{b5} as follows
\be\label{b6}
\bar a_t^{(1)}(Q^2/\Lambda^2) \simeq \frac{1}{\ln(Q^2/\Lambda^2)} - \frac{\pi^2}{3}\frac{1}{\ln^3(Q^2/\Lambda^2)} + \dots ,
\ee
which differs from the exact one-loop result \eqref{eq11}, apart from very small power correction like $\Lambda^2/Q^2$, by terms of order $1/\ln^3(Q^2/\Lambda^2)$. This is a general feature since in the ultraviolet region all running couplings have a similar behaviour, due to asymptotic freedom.

\section{Scheme independence}\label{appC}

As has been mentioned in the Introduction, in Refs. \cite{Deur:2016tte,Stevenson:1980du,Stevenson:1981vj,Grunberg:1980ja,Grunberg:1982fw,Brodsky:2011zza,Brodsky:2011ig,deTeramond:2008ht}, various criteria have been proposed for selecting a renormalization scheme so that a perturbative result best approximates the exact (scheme-independent) result. In this Appendix, rather than discussing any specific choice of renormalization scheme in QCD used to compute $R(s)$ (the cross section for $e^+ e^-\rightarrow$ hadrons) we show how the RG equation allows one to sum all higher order contributions to $R(s)$ that explicitly involve parameters that characterize the renormalization scheme being used, and that this summation leads to cancellation between this explicit dependence on these parameters and implicit dependence on these parameters through the QCD coupling $\alpha$. $R(s)$ is finally expressed as a power series in the
't Hooft coupling evaluated at $s/\Lambda^2$ with coefficients that are independent of the renormaliztion scheme being used.

Let us consider the ratio $R$ which is
a function of the center of mass energy squared $s$ and of the renormalized coupling constant $\alpha$. At high energies, this ratio may be written in terms of a  perturbative series as  
\be\label{eq24a}
R(s/\mu^2,\alpha(\mu^2/\Lambda^2)) =
\sum_i Q_i^2
\left[1+\sum_{n=0}^\infty \sum_{m=0}^n T_{n,m}
\ln^m\left(\frac{s}{\mu^2}\right) \alpha^{n+1}\left(\frac{\mu^2}{\Lambda^2}\right) 
\right],
\ee
where $Q_i\,e$ is the charge on the $i^{th}$ quark. If we group the terms in Eq. \eqref{eq24a} so that
\be\label{e42}
A_n(\alpha) = \sum_{k=0}^\infty T_{n+k,n} \alpha^{n+k+1} 
\ee
then
\be\label{e43}
R = \sum_{i} Q^2_i \sum_{n=0}^\infty A_n(\alpha(\mu^2/\Lambda^2)) \ln^n\left(\frac{s}{\mu^2}\right).
\ee

The RG equation results in $A_{n+1}$ being expressed in terms of $A_n$; when this relation is iterated, $R$ is expressed entirely in terms of $A_0(\alpha)$ with $\alpha$ being evaluated at $s/\Lambda^2$. We then have \cite{r20}
\be\label{e44b}
R(s/\mu^2,\alpha(\mu^2/\Lambda^2))
= \sum_i Q_i^2 A_0(\alpha(s/\Lambda^2))
= \sum_i Q_i^2 \left[1+\sum_{n=0}^\infty\tau_n\alpha^{n+1}(s/\Lambda^2)\right];
\;\;\;\;\; (\tau_n = T_{n,0}).
\ee
As expected, the RG equation leads to an expression for $R$ that is independent of $\mu$.
(This approach can also be used if there are massive fields present \cite{r39}.)
There is no need for devising some criterion for choosing $\mu$ so that when considering a finite number of terms in the series of Eq. \eqref{eq24a}, the perturbative result best approximates the exact result. In Ref. \cite{r20} it is shown that the RG equation allows one to actually sum those terms in $R$ that have an explicit dependence on $\mu$ through $\ln(s/\mu^2)$, and that once this is done, this explicit dependence on $\mu^2$ cancels against the implicit dependence on $\mu^2$ in Eq. \eqref{eq24a} that resides in $\alpha(\mu^2/\Lambda^2)$ (see also \cite{Akrami:2019sru}).

At this stage, the coupling $\alpha$ in Eq. \eqref{eq24a} is dependent on the renormalization scheme used, with the scheme being parametrized by coefficients $\beta_i$ ($i\ge 2$) in Eq. \eqref{eq2}, with $\partial\alpha/\partial\beta_i$ depending on $\beta_i$ \cite{Stevenson:1981vj}. As a result, it is not feasible to effect the sort of summation that leads from Eq. \eqref{eq24a} to Eq. \eqref{e43}, so the explicit and implicit dependence of $R$ on $\beta_i$ cannot be shown to explicitly cancel. However, in Refs. \cite{r21,r22}, it is shown that if one replaces the coupling $\alpha$ by a parameter $z$ which is effectively the 't Hooft coupling, renormalization scheme dependence on the parameters $\beta_i$ ($i\ge 2$) is replaced by dependence on a single parameter $\gamma$. Again, the RG equation can be used to show that all dependence on $\gamma$ cancels in $R$, just as dependence of $R$ on $\mu$ cancels. In addition, the expansion parameters $\tau_n$ in Eq. \eqref{e43} can be explicitly shown to be renormalization scheme independent \cite{r20}. This is why we focus on the 't Hooft coupling when considering the analytic running coupling in section \ref{sec3}.

\end{document}